# 100 years of the first experimental test of General Relativity

Luís C. B. Crispino[1] and Daniel J. Kennefick[2]

*Einstein's general theory of relativity is one of the most important accomplishments in the history of science. Its experimental verification a century ago is therefore an essential milestone that is worth celebrating in full. We reassess the importance of one of the two expeditions that made these measurements possible – a story that involves a sense of adventure and scientific ingenuity in equal measure.*

In the concluding section of their famous paper reporting the successful test of Einstein's General Theory of Relativity (GR) [1], Frank Watson Dyson, Arthur Stanley Eddington and Charles Rundle Davidson noted that "In summarising the results of the two expeditions, the greatest weight must be attached to those obtained with the 4-inch lens at Sobral. From the superiority of the images and the larger scale of the photographs it was recognised that these would prove to be much the most trustworthy."

The test was performed during the total solar eclipse of May 29, 1919 by two British expeditions, one to Príncipe Island, off the western coast of Africa, and the other to the city of Sobral, in North-eastern Brazil.

Eddington and Edwin Turner Cottingham went to Príncipe Island taking with them the object glass of the Oxford astrographic telescope, fed by a 16-inch coelostat – a mirror mounted so as to track with the sky, thereby keeping star images sharp without the need for bulky telescope mounts. Andrew Claude de la Cherois Crommelin and Davidson went to Sobral, taking with them two telescopes (also fed by coelostats), a 13-inch Greenwich astrographic object glass and a back telescope with a 4-inch object glass (cf. Fig. 1).

---

[1] Faculdade de Física, Universidade Federal do Pará, 66075-110, Belém, PA, Brazil.
[2] Department of Physics, University of Arkansas, Fayetteville, AR 72701, USA.

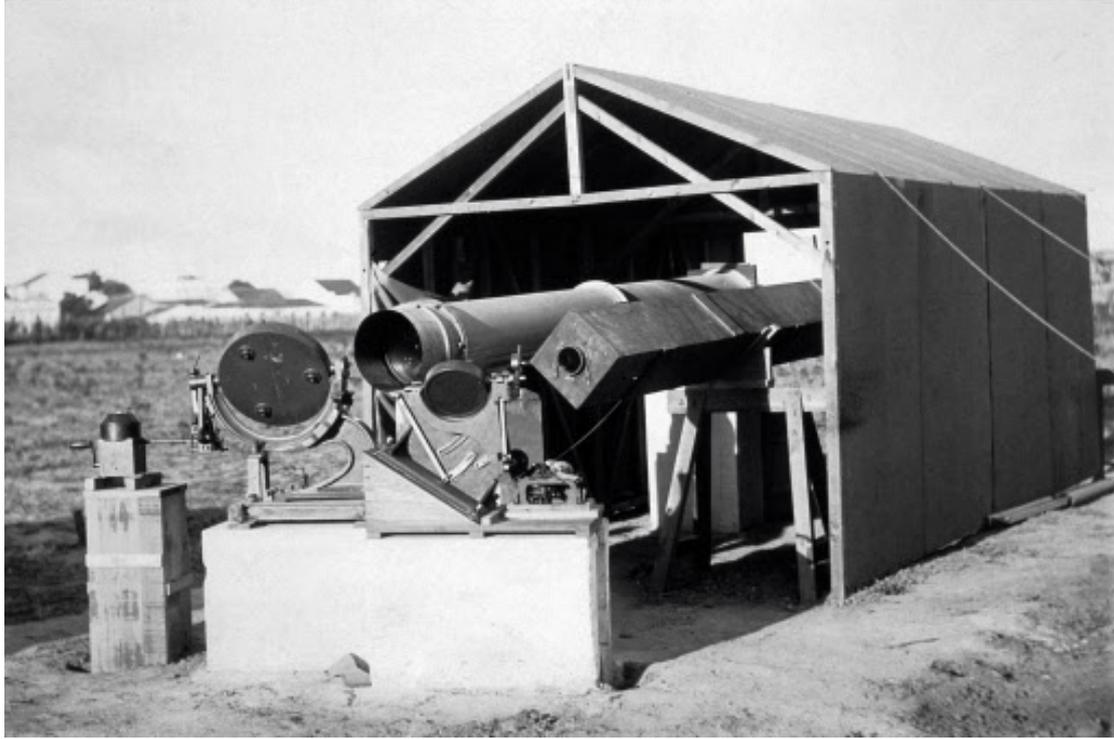

Figure 1 – Telescopes mounted in Sobral, Brazil, in 1919. Courtesy of the Science Museum/Science & Society Picture Library, London, United Kingdom.

The two British commissions left together from Liverpool on the 8[th] of March, on board the steamer *Anselm*. They called at the island of Madeira, where the two teams separated. Eddington and Cottingham spent over a month on the island before obtaining passage on another steamer, the *Portugal*, which disembarked them on Príncipe on April 23. Crommelin and Davidson proceeded aboard the *Anselm* until they reached Belém, in Brazil, where they arrived on March 23. Since they were well in advance of the date of the eclipse, they decided to stay in the *Anselm* during its round trip to Manaus, nearly a thousand miles up the Amazon. In Belém, during their stay, a translation of a paper written by them was published in the local press. Such was the novelty of Einstein's new theory, which was little known abroad because of the wartime break in scientific communications, that this paper was the first printed text dealing with the General Theory of Relativity published in the Americas (cf. Fig. 2) [2]. Crommelin and Davidson then travelled on by boat and train to Sobral, where they were joined by the Brazilian and North-American commissions. The Brazilian team was led by Henrique Morize, the director of the National Observatory, who did much to assist the British expedition's preparations (cf. Fig. 3).

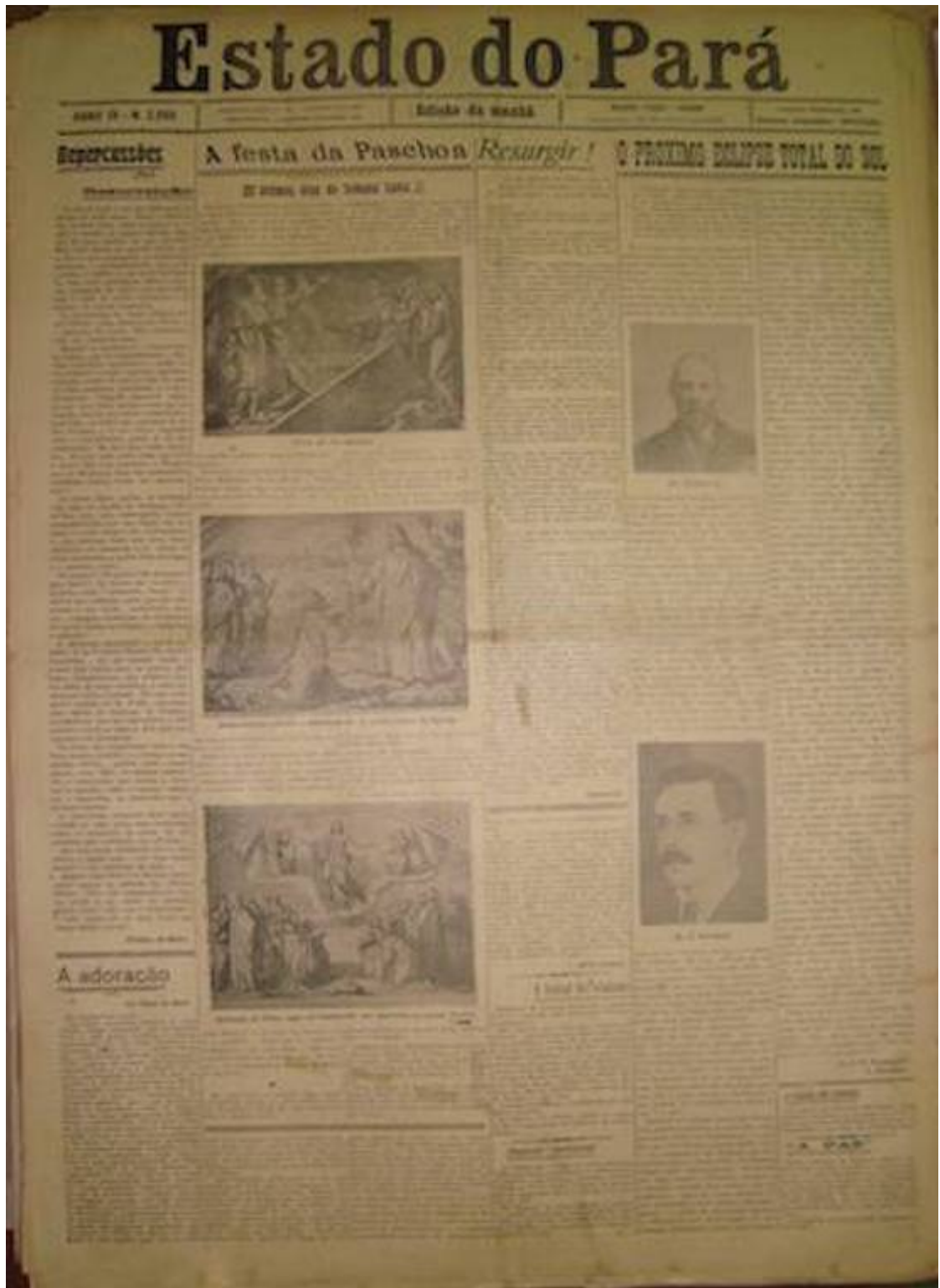

Figure 2 – Cover of the newspaper Estado do Pará, published in Belém, on April 20, 1919, containing a translation of the article signed by Crommelin and Davidson. Courtesy of *Biblioteca Pública Arthur Vianna*, Pará, Brazil.

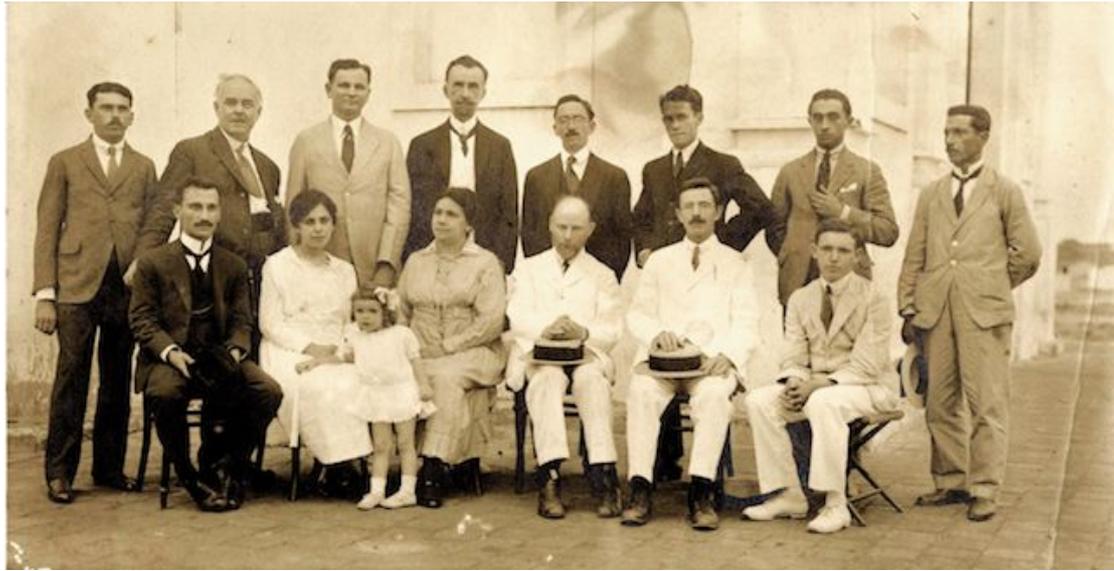
Figure 3 – British, North-American and Brazilian parties at Sobral. Courtesy of *Museu de Astronomia e Ciências Afins*, Rio de Janeiro, Brazil.

On the day of the eclipse unfavourable weather confronted the observers both in Sobral and on Príncipe. At first contact, the beginning of the partial eclipse phase, the Sun was obscured by cloud at both sites. Fortunately for Davidson and Crommelin in Sobral, a few moments before totality, the region of the sky around the Sun cleared and good photographs were obtained by both instruments.

Eddington and Cottingham obtained 16 photographs with the telescope mounted in Príncipe. As Eddington discovered upon developing the plates, the cloud there had thinned towards the end of totality and a handful of stars were imaged on a couple of plates. Hoping to avoid a steamship strike they did not linger on Príncipe, departing on June 12, arriving back in Liverpool on July 14. This meant that they did not take comparison plates on the island. Instead these plates, of the same star field but at night with the Sun absent, had been taken before departure in Oxford with the telescope still mounted in its dome. In order to guard against any change in magnification between the two quite different setups, check plates of a different star field in *Arcturus* were taken both at Príncipe in Oxford.

Crommelin and Davidson obtained 19 photographs with the astrographic object glass and 8 with the 4-inch lens. However, while developing the plates in the days after the eclipse Davidson, as he noted in his diary, discovered that the astrographic telescope had lost focus during the eclipse and that the star images were heavily distorted. His comment at the time was "it seems doubtful whether much can be got from these plates" [3]. Although the data was analysed back at Greenwich, no weight was ultimately assigned to the results obtained from the Sobral astrographic plates.

Comparison photographs of the eclipse star field were taken by Crommelin and Davidson in Sobral in July, before returning to England on August 25. By then Eddington had already conducted an analysis of his data which suggested, as he reported to a meeting of the British Association for the Advancement of Science in September, that the amount of light deflection fell somewhere between the two different predictions made by Einstein. It was news of this presentation, transmitted to Einstein by his Dutch colleague Hendrik A. Lorentz via telegram, which gave rise to the famous Einsteinian quip that he was glad for the Lord's sake that nature had matched up well against his theory. However Eddington's measurements were of low weight, because of the small number of stars he had been able to image. Everything depended upon the analysis of the results from Sobral, conducted at Greenwich under Dyson's direction.

In recent decades commentary on the eclipse of 1919 and its results has focused heavily upon Eddington and his role, including his alleged pro-Einstein bias [4]. The role of the other astronomers involved has been, so to speak, eclipsed by Eddington's (and Einstein's) fame. Dyson anticipated this. He commented to his daughter "if I'm remembered in the future it will be because of my association with Eddington. People will say – Dyson? Oh yes – he was Astronomer Royal, when Eddington was Chief Assistant" [5]. The consequence of the exaggerated focus on Eddington has been an associated tendency to focus on his station at Principe, with Sobral relegated to an afterthought. Because of the problems with the Sobral astrographic many people today imagine that the Sobral expedition played a relatively minor role. In truth, it was central to the success of the whole enterprise.

As presented by Dyson, Eddington and Davidson in their report, the experiment sought to test between three different theoretical predictions. The first was the presumption, inherent in the nineteenth-century wave theory of light, that light has no mass and is unaffected by gravity. As such the presence of the Sun would cause no deflection of stars in its field. The second possibility was put forward by Einstein as a consequence of his principle of equivalence. In this viewpoint light has energy, which means it has mass. Thus it falls towards the Sun as it passes by, causing a small deflection (0.87 arcsecond at the limb of the Sun) in star positions away from the Sun, as seen from Earth. Finally, after developing GR, with its prediction that gravity alters the geometry of spacetime, Einstein realized there would be an additional deflection, due to curvature near the Sun. This resulted in his final light deflection prediction (1.75 arcsecond at the limb of the Sun), twice as great as his original one. In their presentations, Eddington and Dyson chose to assign credit for the middle ("half-deflection") prediction, to Newton, on the grounds that it was consistent with massive photons interacting with the Sun according to his famous law of gravity [6].

The results confirming Einstein's theory were announced during a joint meeting of the Royal Society and the Royal Astronomical Society, held in November 6, 1919 at Burlington House, Piccadilly, London. Newspapers and magazines all around the globe took note of the announcement, making Einstein's name and theory famous worldwide (cf. Fig. 4).

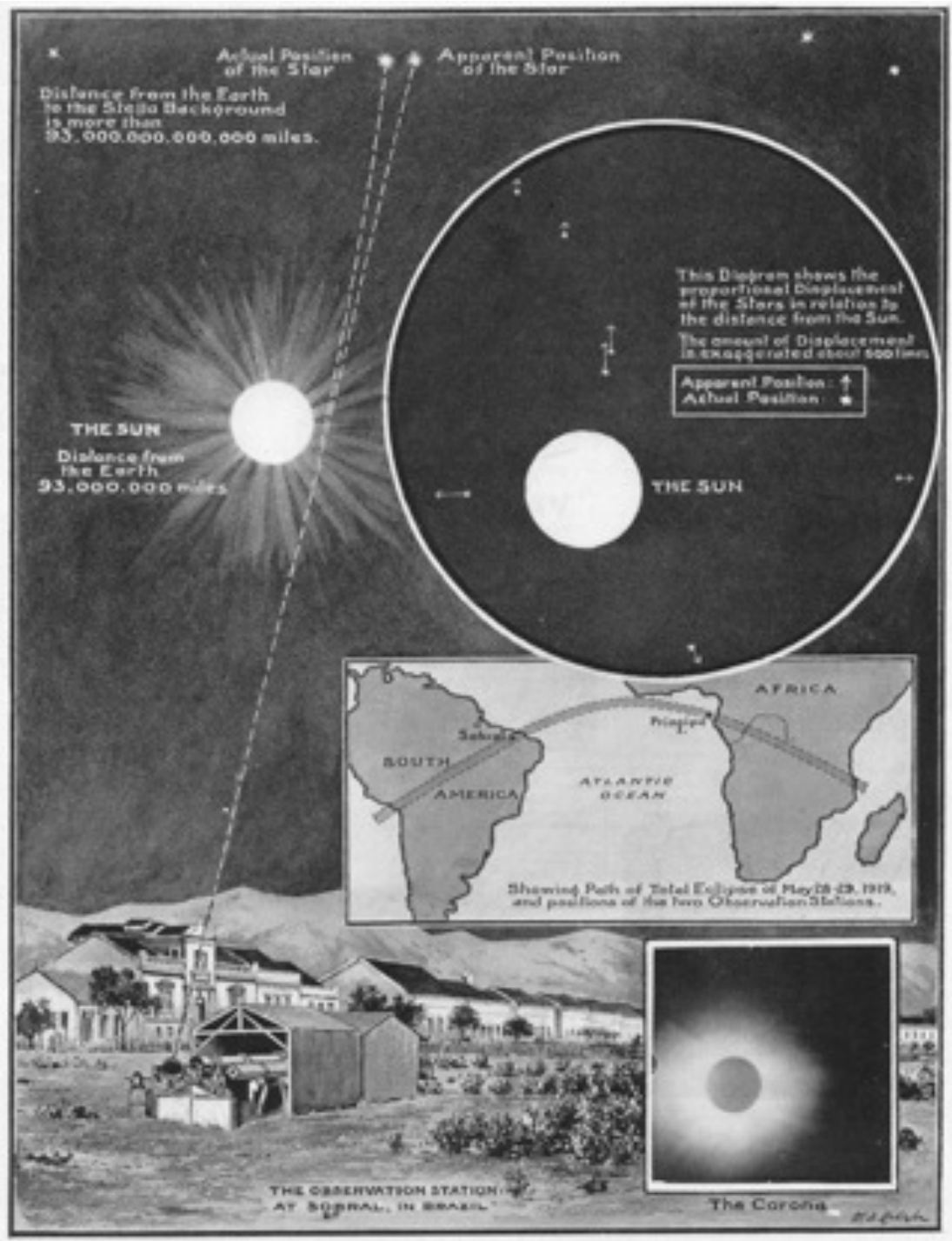

Figure 4 – Page of the November 22, 1919 edition of The Illustrated London News. Courtesy from Illustrated London News, London, United Kingdom.

The expeditions' results received careful scrutiny from sceptics of GR for years after the announcement. But in 1922, and at subsequent eclipses, similar measurements of starlight deflections confirmed General Relativity's triumph over Newton's theory. But in recent decades doubts concerning the soundness of the conclusions of the British experiments in 1919 that have been expressed by some physicists and historians of science [7]. These modern doubts have principally concerned the decision to throw out the data from the Sobral astrographic. Allegations have been raised that Eddington's favourably disposed attitude to GR was responsible for this decision.

The plates obtained with the three telescopes led to the following results (cf. Table). The Sobral Astrographic's result was close to the so-called "Newtonian" result (the "half-deflection"). Eddington's Principe data obtained a result between that and the full GR deflection, but much closer to the latter. The four-inch from Sobral's result was somewhat greater than the GR prediction. The claim that Eddington acted wrongly in throwing out the Sobral astrographic data fails on a number of grounds [7]. First of all the decision to throw out this data was taken by Dyson, in consultation with Davidson, and not by Eddington. Dyson did not share Eddington's bias in favour of GR. Secondly a study of Dyson's data analysis sheets show that he and his team analysed their astrographic data to show that the agreement with Newton could only hold if the instrument had undergone a large change of magnification, in addition to its loss of focus during the eclipse [8]. In other words, only if the astrographic had malfunctioned could the data confirm Newton's Law. If the magnification was presumed not to have changed, they calculated that the instrument would have agreed with Einstein and the two other telescopes (cf. Table). In addition they relied on Davidson's judgement, made before data analysis began, that the instrument, which he operated during the eclipse, was not trustworthy.

One overlooked modern contribution to the debate was provided by the astrometry team at the Royal Greenwich Observatory (RGO) in the late seventies [9]. Inspired by the centenary of Einstein's birth in 1979 they took out the Sobral plates from both telescopes used there and remeasured them using a modern plate measuring machine (cf. Table). They reduced the data with astrometric data reduction software on electronic computers. In 1919 the principal computer had been Davidson, who began his working life at Greenwich in the 19$^{th}$ century, when a computer was a person hired to work problems by hand. This team, led by Andrew Murray, not only vindicated the result, and its claimed error, from the four-inch lens, but also found a result for the astrographic lens in close agreement with the alternative value calculated by Dyson and his team in their notes (and mentioned briefly in their report). Thus this modern re-analysis provides ample justification for the key decision taken by Dyson to reject the apparent confirmation of Newton by the first analysis of the astrographic data.

| Instrument | 1919 Result [1] | 1979 Re-analysis [9] |
|---|---|---|
| Príncipe astrographic | 1.61 ± 0.30 | --- |
| Sobral 4 inch | 1.98 ± 0.18 | 1.90 ± 0.11 |
| Sobral astrographic | 0.93 ± 0.50 or 1.52 ± 0.46 | 1.55 ± 0.34 |

Table – Experimental measurements (in arcsecond) obtained from the 1919 plates. The instruments are listed in the left column. The central column shows the results obtained in 1919 [1], including results from two different calculations based upon the Sobral astrographic data, and the results obtained from the remeasurement of the Sobral plates, carried out later at the RGO [9], are displayed in the right column.

Unfortunately, the focus in recent accounts on the controversy over the Sobral astrographic data has spread the impression amongst some readers that no reliable data was taken at Sobral and that the experimental verdict belonged to Eddington at Principe. This impression is quite mistaken. Eddington suffered, through no fault of his own, from a paucity of data. It was only the superb quality of the images taken at Sobral with the four-inch instrument, operated by Crommelin and overhauled by Davidson, that permitted the celebrated decision in favour of Einstein to be made. In this year of the centenary, the contribution of both of these men, and Dyson, should be restored to their rightful place alongside Eddington in this story of great scientific enterprise.

100 years later, General Relativity lives through another triumphant era, brought about by the detection of gravitational waves (GW) by the LIGO/VIRGO collaboration. This has brought to light the existence of black holes weighing tens of solar masses and given birth to a new era of multi-messenger astronomy. More discoveries doubtless await as the existing detectors continue to improve their sensitivity and look forward to a second century of relativity, as exciting as the first.

For the centenary of this historical eclipse, celebrations are planned. The Portuguese Society of Relativity and Gravitation is organizing a conference in Príncipe Island, and the Brazilian Society for the Progress of Science is hosting a meeting in Sobral. With luck the occasion of the centenary will draw attention to the important role played by the Greenwich team who travelled to Sobral one hundred years ago, amidst many difficulties, to accomplish one of the greatest experiments in the history of science.